\documentclass[10pt]{llncs}
\newcommand{\ignore}[1]{}
\usepackage[pass]{geometry}
\usepackage{fancyhdr}
\usepackage[normalem]{ulem}
\usepackage[hyphens]{url}
\usepackage{hyperref}
\usepackage{color}
\usepackage[english]{babel}
\usepackage{caption}
\usepackage{enumitem}
\setenumerate[1]{itemsep=0pt,partopsep=0pt,parsep=\parskip,topsep=5pt}
\setitemize[1]{itemsep=0pt,partopsep=0pt,parsep=\parskip,topsep=5pt}
\setdescription{itemsep=0pt,partopsep=0pt,parsep=\parskip,topsep=5pt}
\usepackage{booktabs}

\usepackage{cite}
\usepackage{amsmath,amssymb,amsfonts}
\usepackage{algorithmic}
\usepackage{graphicx}
\usepackage{textcomp}
\usepackage{xcolor}

\fancypagestyle{firstpage}{
  \fancyhf{}
\setlength{\headheight}{50pt}

  \pagenumbering{arabic}
}

\title{Zipper Stack: Shadow Stacks Without Shadow\thanks{This work is partially supported by the National Natural Science Foundation of China (No. 61602469, U1836211), and the Fundamental theory and cutting edge technology Research Program of Institute of Information Engineering, CAS(Grant No. Y7Z0411105).}}
\author{Jinfeng Li\and Liwei Chen\thanks{Corresponding Author}\and Qizhen Xu\and Linan Tian\and Gang Shi\and Kai Chen\and \\ Dan Meng}
\institute{Institute of Information Engineering, Chinese Academy of Sciences \and School of Cyber Security, University of Chinese Academy of Sciences \email{\{lijinfeng,chenliwei,xuqizhen,tianlinan,shigang,chenkai,mengdan\}@iie.ac.cn}}

\begin{document}
\maketitle
\pagestyle{plain}


\begin{abstract}

 Return-Oriented Programming (ROP) is a typical attack technique that exploits return addresses to abuse existing code repeatedly. Most of the current return address protecting mechanisms (also known as the Backward-Edge Control-Flow Integrity) work only in limited threat models. For example, the attacker cannot break memory isolation, or the attacker has no knowledge of a secret key or random values. 

 This paper presents a novel, lightweight mechanism protecting return addresses, Zipper Stack, which authenticates all return addresses by a chain structure using cryptographic message authentication codes (MACs). This innovative design can defend against the most powerful attackers who have full control over the program's memory and even know the secret key of the MAC function. This threat model is stronger than the one used in related work. At the same time, it produces low-performance overhead. We implemented Zipper Stack by extending the RISC-V instruction set architecture, and the evaluation on FPGA shows that the performance overhead of Zipper Stack is only 1.86\%. Thus, we think Zipper Stack is suitable for actual deployment.
\end{abstract}


\section{Introduction}

In the exploitation of memory corruption bugs, the return address is one of the most widely exploited vulnerable points. On the one hand, code-reuse attacks (CRAs), such as ROP \cite{ROP-RISC:2008} and ret2libc \cite{return-into-libc:2007}, perform malicious behavior by chaining short sequences of instructions which end with a return via corrupted return addresses. These attacks require no code injection so they can bypass non-executable memory protection. On the other hand, the most widely exploited memory vulnerability, stack overflow, is also exploited by overwriting the return address. Both CRAs and stack smashing attacks are based on tampering with the return addresses.

In order to protect the return addresses, quite a few methods were presented, such as Stack Protector (also known as Stack Canary) \cite{stackguard:1998,Stackguard:2003}, Address Space Layout Randomization (ASLR) \cite{ASLR:2003}, Shadow Stacks \cite{SS:2001,SS:2008,Cost-of-SS:2015,SSbinary:2003}, Control Flow Integrity (CFI) \cite{CFI:2005,CFI2:2009}, and Cryptography-based CFI \cite{CCFI:2015,Pointer-Authentication}. However, they have encountered various problems in the actual deployment.

Stack Protector and ASLR rely on secret random values (cookies or memory layout). Both methods are widely deployed now. However, if there is a memory leak, both methods will fail \cite{Break-memory:2009}: the attacker can read the cookie and hold it unchanged while overwriting the stack to bypass the Stack Protector, and de-randomize ASLR to bypass ASLR. Some works have proved that they can be reliably bypassed in some circumstances \cite{bypassASLR,exploitASLR,Side-Channel_ASLR}, such as BROP \cite{BROP}. Even if there is no information leaking, some approaches can still bypass ASLR and perform CRAs \cite{Seibert:2014}.

Shadow Stack is a direct mechanism that records all return addresses in a protected stack and checks them when returns occur. It has been implemented via both compiler-based and instrumentation-based approaches \cite{ROPdefender:2011,Cost-of-SS:2015}. In recent years, commercial hardware support has also emerged \cite{cet2016}. But Shadow Stack relies heavily on the security of the memory isolation, which is difficult to guarantee in actual deployment. 
Some designs of Shadow Stack utilize ASLR to protect Shadow Stack. However, they cannot thwart the attacks contains any information disclosure \cite{Cost-of-SS:2015, losing}. Since most methods that bypass ASLR \cite{bypassASLR,exploitASLR} are effective against this type of defense. Other designs use page attributes to protect Shadow Stack, for instance, CET \cite{cet2016}. However, defenses that rely on page attributes, such as NX (no-execute bit), have been bypassed by various technologies in actual deployment \cite{bypass-dep}: a single corrupted code pointer to the function in the library (via a JOP/COOP attack) may change the page attribute and disable the protection. 
In light of these findings, we think that mechanisms that do not rely on memory isolation are more reliable and imperative.

Cryptography-based protection mechanisms based on MAC authentication have also been proposed. These methods calculate the MACs of the return addresses and authenticate them before returns \cite{CCFI:2015,RAGuard:2017,Pointer-Authentication}. They do not rely on memory isolation because the attacker cannot generate new correct MACs without the secret key. But they face other problems: replay attacks (which reuse the existing code pointers and the MACs of them), high-performance overhead, and security risk of keys. A simple authentication of MAC is vulnerable facing replay attacks, so some designs introduce extra complexity to mitigate the problem (such as adding a nonce or adding stack pointer into the MAC input). Besides, these methods' performance overhead is also huge, since they use a cryptographic MAC and then save the result into memory. Both operations consume massive runtime overhead. The security of the secret key is also crucial. Theses works assume that there is no hardware attack or secret leak from the kernel. However, in the real world, the secret key can be leaked by a side-channel attack or an attack on context switching in the kernel. Therefore, we should go further over these assumptions, and consider a more powerful threat model.

\begin{figure}[hbt]
\begin{center}
\includegraphics[width=0.5\textwidth]{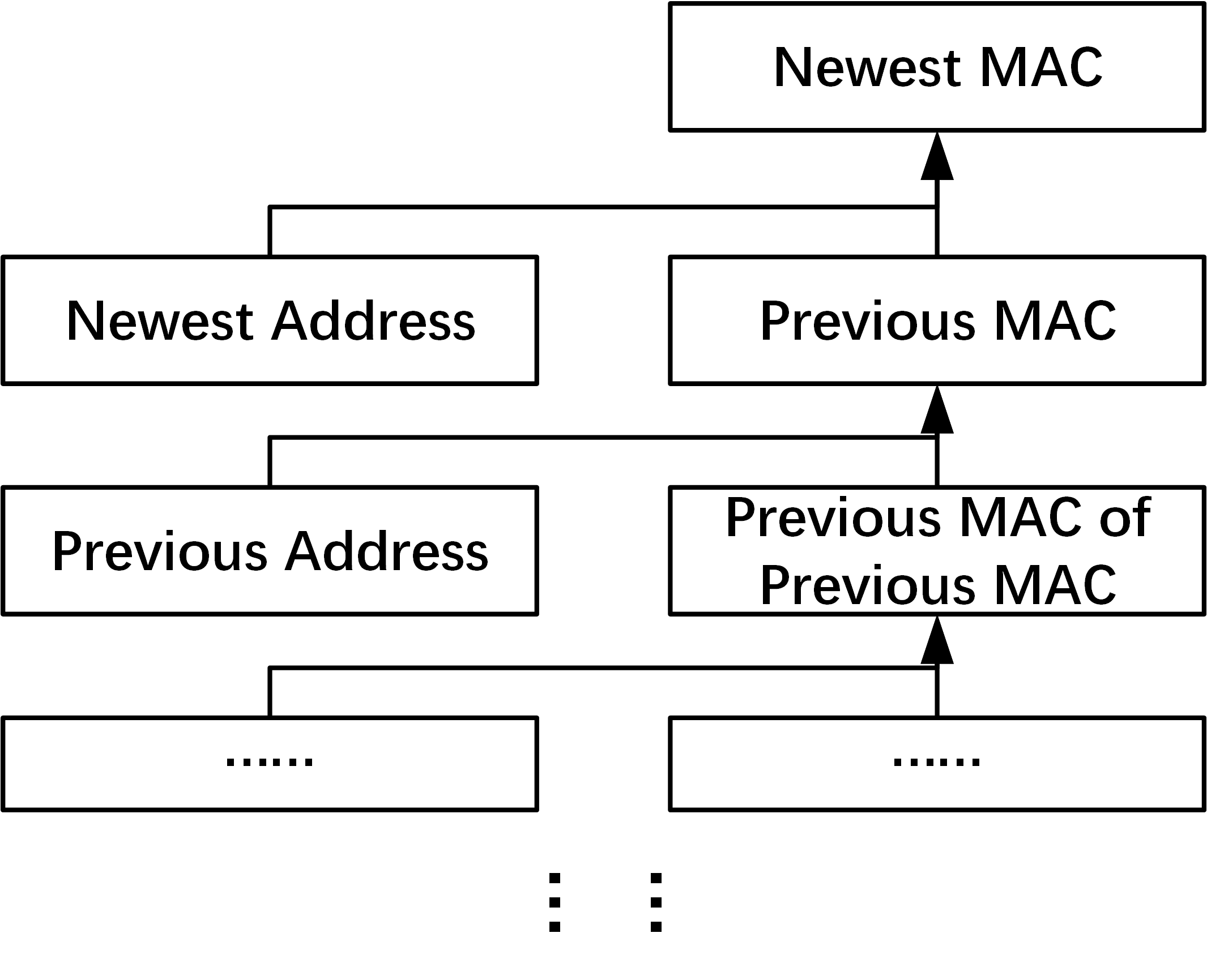}
\end{center}
\vspace{-1.5em}
\caption{\small{Core of Zipper Stack}}
\label{intro}
\end{figure}

In this paper, we propose a novel method to protect return addresses, Zipper Stack, which uses a chain of MACs to protect the return addresses and all the MACs. In the MAC chain, the newest MAC is generated from the latest return address and the previous MAC. The previous MAC is generated from the return address and MAC before the previous one, as Figure \ref{intro} shows. So the newest MAC is calculated from all the former return addresses in the stack, although it is generated by computing the latest address and the previous MAC. Zipper Stack minimizes the amount of state requiring direct protection: only the newest MAC needs to be protected from tampering. Without tampering the newest MAC, an attacker cannot tamper with any return address because he cannot tamper the whole chain and keep the relation.

\vspace{-0.5em}

Zipper Stack avoids the problems of Shadow Stack and Cryptography-based protection mechanisms. Compared to Shadow Stack, it does not rely on memory isolation. Consequently, the attacks that modify both the shadow stack and the main stack cannot work in Zipper Stack. Compared with other cryptography-based mechanisms, Zipper Stack can resist against replay attacks itself and will not fail even if the secret key is leaked (see Section \ref{s_design}). In terms of efficiency, Zipper Stack performs even better. Our design is more suitable for parallel processing in the CPU pipeline, which avoids most performance overhead caused by the MAC calculation and memory access. The performance overhead of Zipper Stack with hardware support based on Rocket Core \cite{rocket} (RISC-V CPU \cite{riscv}) is only 1.86\% based on our experiments.

The design of Zipper Stack solved three challenges: First, it avoids the significant runtime overhead that most cryptography-based mechanisms suffer since the newest MAC is updated in parallel. In our hardware implementation, most instructions that contain a MAC generation/authentication take only one cycle (See Section VI). Second, it utilized the LIFO order of return addresses to minimize the amount of state requiring direct protection. In general, a trust root authenticating all the data can help us defend against replay attacks or attacks containing secret key leaks (which most current return address protection method cannot). While in Zipper Stack, the authentication uses the newest MAC, at the same time, the MAC is a dynamic trust root itself. So it gets better security without extra overhead. Third, previous methods protect each return address separately, so any one been attacked may cause attacks. Zipper Stack, however, connect all the return addresses, leverage the prior information to increase the bar for attackers.



In order to demonstrate the design and evaluate the performance, we implement Zipper Stack in three deployments corresponding to three situations: 

a) Hardware approach, which is suitable when hardware support of Zipper Stack is available.
b) Customized compiler approach, which is suitable when hardware support is not available, but we can recompile the applications.
c) Customized ISA approach, which is suitable when we cannot recompile the programs, but we can alter the function of CALL/RET instructions.

Ideally, the hardware approach is the best - it costs the lowest runtime overhead. The other two approaches, however, are suitable in some compromised situation. In the hardware approach, we instantiated Zipper Stack with a customized Rocket Core (a RISC-V CPU) on the Xilinx Zynq VC707 evaluation board (and hardware-based Shadow Stack as a comparison). In customized compiler approach, we implemented Zipper Stack in LLVM. In customized ISA approach, we use Qemu to simulate the modified ISA.

\textbf{Contributions.} In summary, this paper makes the following contributions:
\begin{enumerate}
\item \textbf{Design:} We present a novel, concise, efficient return address protection mechanism, called Zipper Stack, which protect return addresses against the attackers have full control of all the memory and know the secret keys, with no significant runtime overhead. Consequently, we analyze the security of our mechanism.
\item \textbf{Implementation:} To demonstrate the benefits of Zipper Stack, we implemented Zipper Stack on the FPGA board, and a hardware-based Shadow Stack as a comparison. To illustrate the potential of Zipper Stack to be further implemented, we also implemented it in LLVM and Qemu.
\item \textbf{Evaluation:} We quantitatively evaluated the runtime performance overhead of Zipper Stack, which is better than existing mechanisms.
\end{enumerate}

\section{Background and Related Work}
\label{s_background}

\subsection{ROP attacks}
Return Oriented Programming (ROP) \cite{return-into-libc:2007,ROP-RISC:2008} is the major form of code reuse attacks. ROP makes use of existing code snippets ending with return instructions called gadgets to perform malicious acts. In ROP attacks, the attackers link different gadgets by tampering with a series of return addresses. An ROP attack is usually made up of multiple gadgets. At the end of each gadget, a return instruction links the next gadget via the next address in the stack. The defenses against ROP mainly prevent return instructions from using corrupted return addresses or randomize the layout of the codes.

\subsection{Shadow Stack and SafeStack}

Shadow Stack is a typical technique to protect return addresses. Shadow Stack saves the return addresses in a separate memory area and checks the return addresses in the main stack when returns. It has been implemented in both compiler-based and instrumentation-based approaches \cite{SS:2001,SS:2008,Cost-of-SS:2015,smashguard:2006,stackghost:2001}. SafeStack \cite{safestack} is a similar way, which moves all the return addresses into a separated stack instead of backs up the return addresses. SafeStack is now implemented in LLVM as a component of CPI \cite{CPIinLLVM}. 

An isolated stack mainly brings about two problems: One is that memory isolation costs more memory overhead and implementation complexity. Another one is that security relies on the security of memory isolation, which is impractical. As the structure of the shadow stack is simply the copies of return addresses, it is fragile once the attacker can modify its memory area. For example, in Intel CET, the shadow stack's protection is provided by a new page attribute. But a similar approach in DEP is easily bypassed by a variety of methods modifying the page attribute in real-world attacks \cite{bypass-dep}. ASLR is also bypassed in real-world attacks that other implementations rely on. The previous work \cite{losing} constructed a variety of attacks on Shadow Stack.



\subsection{CFI and Crypto-based CFI}
Control Flow Integrity (CFI), which first introduced by Abadi et. al. \cite{CFI:2005, CFI2:2009}, has been recognized as an important low-level security property. In CFI, runtime checks are added to enforce that jumps and calls/rets land only to valid locations that have been analyzed and determined ahead of execution \cite{CFI-rand:2013,Forward-Edge-CFI:2014,Toughcall:2016}. The security and performance overhead of different implementations differ. Fine-grained CFI approaches will introduce significant overhead. However, coarse-grained CFI has lower performance overhead but enforces weaker restrictions, which is not secure enough \cite{losing}. Besides, Control-Flow Graphs (CFGs), which fine-grained CFI bases on, are constructed by analyzing either the disassembled binary or source code. CFG cannot be both sound and complete, so even if efficiency losses are not mainly considered, CFI is not a panacea for code reuse attacks \cite{Control-Jujutsu:2015}. Due to the above reasons, CFI is not widely deployed on real systems now.

To improve CFI, some implementations also introduce cryptography methods to solve problems such as inaccurate static analysis: CCFI \cite{CCFI:2015}, RAGuard \cite{RAGuard:2017}, Pointer-Authentication\cite{Pointer-Authentication}. Most of these methods are based on MAC: the MACs of protected key pointers, including return addresses, are generated, whereafter, authenticated before use. But all of these methods rely heavily on secret keys and cost tremendous performance overhead. 
Another problem of these mechanisms is replay attacks. The attackers can perform replay attacks by reusing the existing values in memory.




\section{Threat Model}
\label{s_threatmodel}

In this paper, we assume that a powerful attacker has the ability to read and overwrite arbitrary areas of memory. He tries to perform ROP (or ret2lib) attacks. This situation is widespread - for example, a controllable pointer out of bounds can help the attacker acquire the capability. Reasonably, the attacker cannot alter the value in the dedicated registers (called Top and Key registers in our design), since these registers cannot be accessed by general instructions.

The attacker in our assumption is more powerful than all previous works. Shadow Stacks assume that the attackers cannot locate or overwrite the shadow stack, which is part of the memory. In our work, we do not need that assumption, which means it can defend against more powerful attacks.

\section{Design}
\label{s_design}

In this section, we elaborate on the design of Zipper Stack in detail. Here, we take the hardware approach as an example. The design in other approaches is equivalent.

\subsection{Overview}

\begin{figure}[hbt]
\begin{center}
\includegraphics[width=0.5\columnwidth]{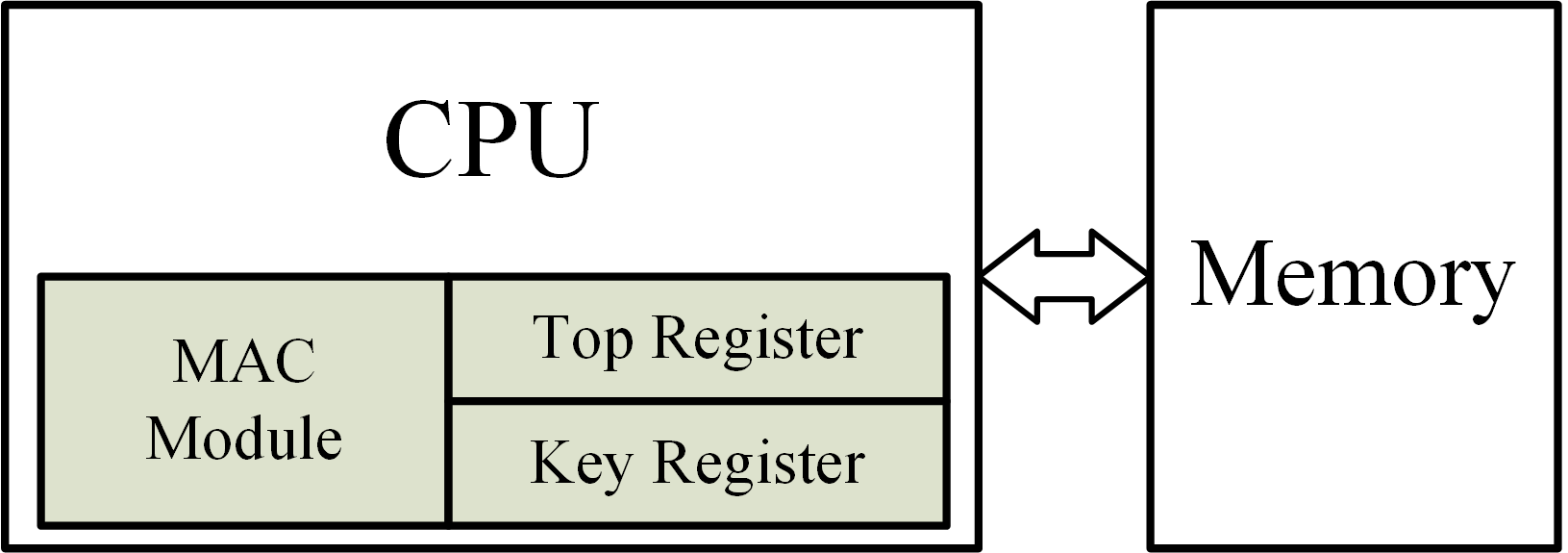}
\end{center}
\vspace{-1.5em}
\caption{\small{Overview of Zipper Stack (hardware approach)}}
\label{hardware}
\end{figure}
\vspace{-1.5em}

In the hardware approach, we need hardware support and the modification of memory layout. Figure \ref{hardware} shows the overview of the hardware in Zipper Stack: Zipper Stack needs two dedicated registers and a MAC module in the CPU, but it requires no hardware modification of the memory. 
The registers include the Top register holding a MAC ($Nm$ bits) and the Key register holding a secret key ($Ns$ bits). 
Both the registers are initialized as random numbers at the beginning of a process, and they cannot be read nor rewritten by attackers. 
The secret key will not be altered in the same process. 
Therefore, we temporarily ignored this register for the sake of simplicity in the following. 
Assuming that the width of return addresses is $Na$, the MAC module should perform a cryptographic MAC function with an input bit width of $Na + Nm$ and an output bit width of $Nm$.

\begin{figure}[hbt]
\begin{center}
\includegraphics[width=0.6\textwidth]{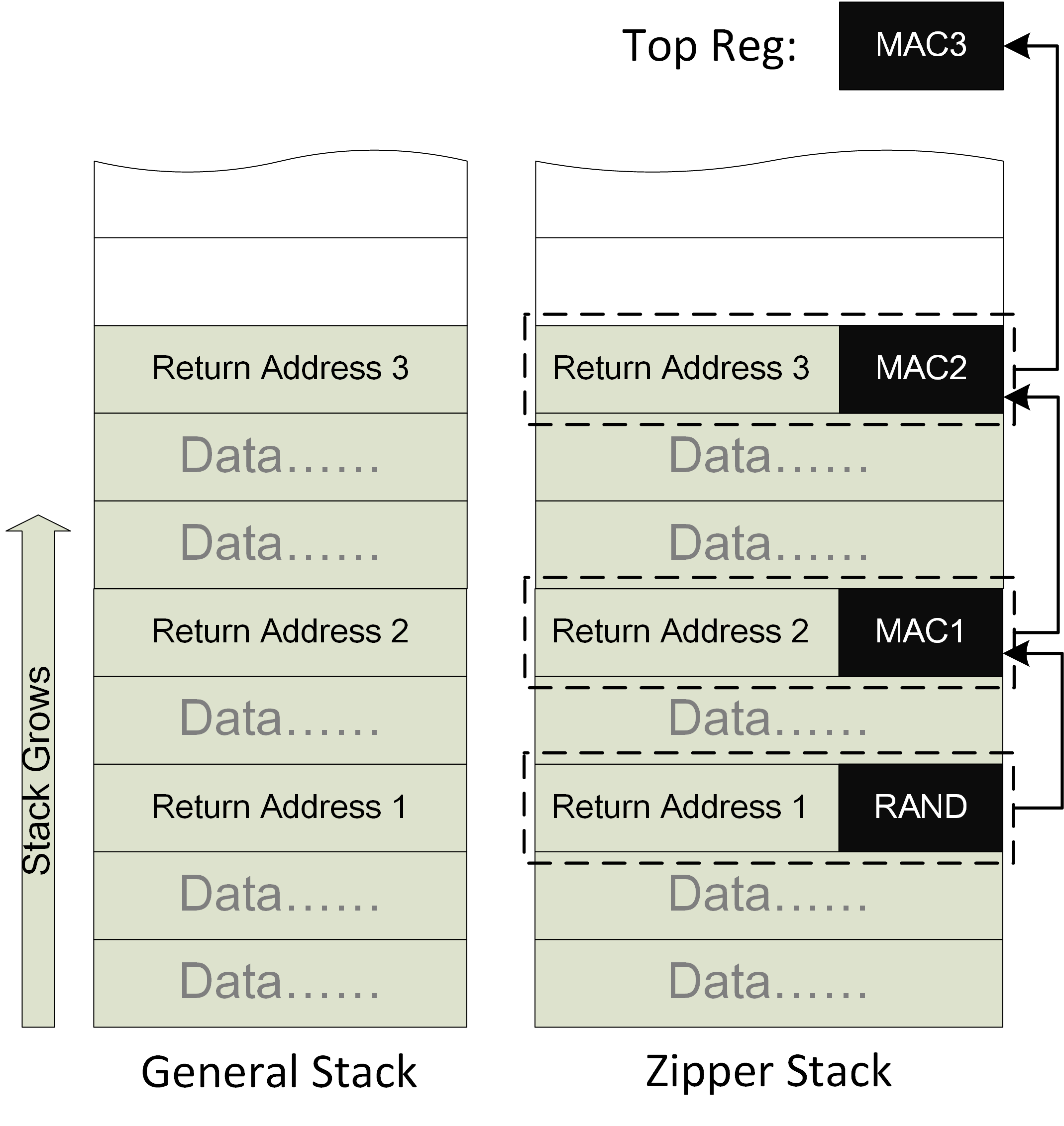}
\end{center}
\vspace{-1.5em}
\caption{\small{Layout in Zipper Stack: The dotted rectangles in the figure indicate the input of the MAC function, and the solid lines indicate the storage location of the MAC.}}
\label{zipper}
\end{figure}

We now turn to the memory layout of Zipper Stack. In Zipper Stack, all return addresses are bound to a MAC, as shown in Figure \ref{zipper}. The novelty is, the MAC is not generated from the address bound with itself, but from the previous return address with the MAC bound with that address. This connection keeps all return addresses and MAC in a chain. To maintain the structure, the top one, namely the last return address pushed into the stack, is handled together with the previous MAC and the new MAC is saved into the Top register; while the bottom one, i.e., the first return address pushed into the stack is bound to a random number (exactly the initial value of Top register when the program begins).

\subsection{Operations}

\begin{figure}[!hbt]
\begin{center}
\includegraphics[width=\textwidth]{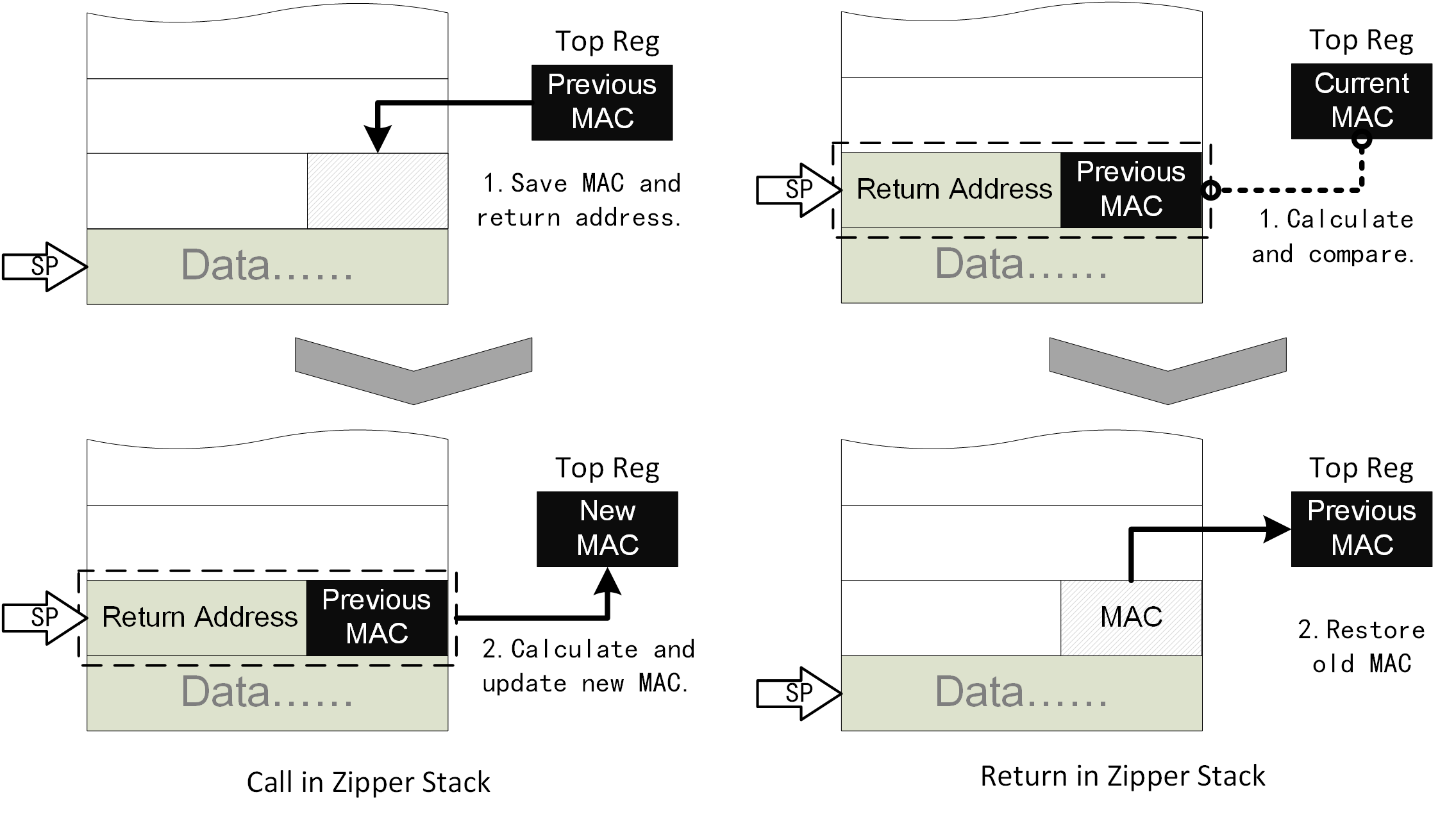}
\caption{\small{The stack layout before/after a call/return. Previous MAC is generated from previous return address and the MAC with that return address. SP stands for stack pointer.}}
\label{callret}
\end{center}
\vspace{-2em}
\end{figure}

Next, we describe how Zipper Stack works with return addresses in the runtime, i.e., how to handle the Call instructions and the Return instructions. As Figure \ref{callret} shows.

\textbf{Call:} In general, the Call instructions perform two operations. First, push the address of the next instruction into the stack. Then, set the PC to the call destination. While in Zipper Stack, the Call instructions become slightly more complicated and need three steps:

\begin{enumerate}
\item Push the Top register along with the return address into the main stack;
\item Calculate a new MAC of the Top register and the return address and save the new MAC into the Top register;
\item Set the PC to the call destination.
\end{enumerate}

\textbf{Return:} In general, the Return instructions also perform two operations. First, pop the return address from the stack; second, set the PC to this address. Correspondingly, in Zipper Stack, Returns also become a little more complicated, including three steps.

\begin{enumerate}
\item Pop the return address and the previous MAC from the main stack and calculate the MAC of them. Then check whether it matches the Top register. If not, raise an exception (which means an attack).
\item Save the MAC poped from the stack into the Top register.
\item Set the PC to the return address.
\end{enumerate}

Figure \ref{callret} shows the process of CALL and RET in Zipper Stack. We omit the normal operations about the PC and return addresses.

The core idea of Zipper Stack is to use a chain structure to link all return addresses together. Based on this structure, we only need to focus on the protection and verification of the top of the chain instead of protecting the entire structure. Just like a zipper, only the slider is active, and when the zipper is pulled up, the following structure automatically bites up. Obviously, protecting a MAC from tampering is much easier than protecting a series of MAC from tampering: Adding a special register in the CPU is enough, and there is no need to protect a special memory area.

\subsection{Setjump/Longjump and C++ Exceptions}
In most cases, the return addresses are used in a LIFO order (last in, first out). But there are exceptions, such as setjump/longjump and C++ exceptions. 
Consequently, most mechanisms protecting return addresses suffer from the Setjump/Longjump and C++ Exceptions, some papers even think the block-chaining like algorithm cannot work with exceptional control-flows \cite{stackghost:2001}. However, 
Zipper Stack can accommodate both Setjump/Longjump and C++ Exceptions. Both Setjump/Longjump and C++ Exceptions mainly save and restore the context between different functions. The main task of them is stack unwind. 
The return addresses in the stack will not encounter any problem since Zipper Stack does not alter either the value nor the position of return addresses. The only problem is how to restore the Top register. The solution is quite simple: backup the Top register just like backup the stack pointer or other registers (in Setjump/Longjump save them in the jump buffer, similar in C++ Exceptions). When we need to Longjump or handle an exception, restore the Top register just as restoring other registers. The chain structure will remain tight. 

However, the jump buffer is in memory, so this solution exposes the Top register (since the attacker can write on arbitrary areas of memory) and leaves an opportunity to overwrite the Top register. So additional protection is a must. In our implementation, we use a MAC to authenticate the jump buffer (or context record in C++ exception). In this way, we can reuse the MAC module. 

\section{Implementations}
\label{s_implementation}

\subsection{Hardware approach}

We introduce the hardware approach first. In hardware approach, we implemented a prototype of Zipper Stack by modifying the Rocket Chip Generator \cite{rocket} and customized the RISC-V instruction set accordingly. We also added a MAC module, several registers, and several instructions into the core. Whereafter, we modified the toolchain, including the compiler and the library glibc. Besides, we implemented a similar Shadow Stack for comparison.

\subsubsection{Core and New Instructions}

In Rocket core, we added a Top register and a Key register, which correspond to those designed in the algorithm. These two registers cannot be loaded/stored via normal load/store instructions. At the beginning of a program, the Key and Top register are initialized by random values.

In RISC-V architecture, a CALL instruction will store the next PC, i.e., return address, to the \emph{ra} register, and a RET instruction will read the address in the \emph{ra} register and jump to the address. Consequently, two instructions were added in our prototype: ZIP (after call), UNZIP (before return). They will perform as a Zipper Stack's CALL/RET together with a normal CALL/RET.

For the sake of simplicity, we use a compressed structure. In RISC-V architecture, the return address in register \emph{ra}, so we put the return address and the MAC together into \emph{ra}. In the current Rocket core, only lower 40 bits are used to store the address. Therefore, we use the upper 24 bits to hold the MAC. Correspondingly, our Top register is 24 bits. And our Key register is 64 bits.

Our new instructions will update a MAC (ZIP after a CALL) or check and restore a MAC (UNZIP before a RET). When a ZIP instruction is executed, the address in \emph{ra} (only lower 40 bits) along with the old MAC in the Top register will be calculated into a new MAC. The new MAC is stored in the Top register. The old MAC is stored to the higher 24 bits in the \emph{ra} register (the lower 40 bits remain unchanged). Correspondingly, when an UNZIP instruction is executed, the \emph{ra} register (including the MAC and address) is calculated and compared with the Top register. If the values match, the MAC in \emph{ra} (higher 24 bits) is restored into the Top register, and the higher 24 bits in \emph{ra} is restored to zero. If the values do not match, an exception will be raised (which means an attack). 


\subsubsection{MAC Module}

Next, we added a MAC module in the Rocket Core. Here, we use Keccak \cite{keccak} (Secure Hash Algorithm 3 (SHA-3) in one special case of Keccak) as the MAC function. In our hardware implementation, the arguments of Keccak are as follows: $l=4, r=256, c=144$ The main difference between our implementation and SHA-3 is that we use a smaller $l$: in SHA-3, $l=6$. 
This module will take 20 cycles for one MAC calculation normally, and it costs 793 LUTs and 432 flip flops. 


\subsubsection{Pipeline}

\vspace{-1em}
\begin{figure}[hbt]
\begin{center}
\includegraphics[width=0.65\columnwidth]{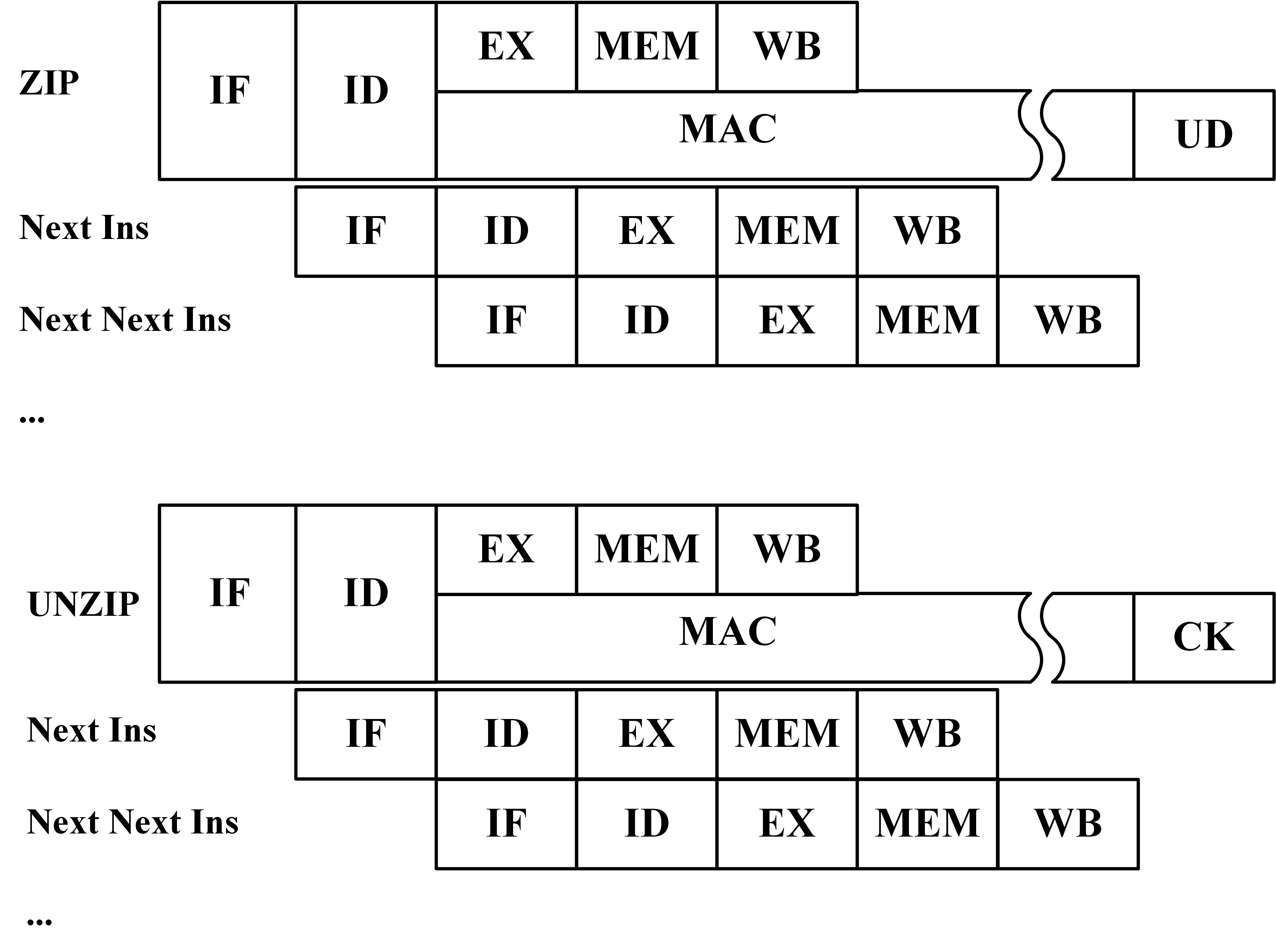}
\end{center}
\vspace{-1.5em}
\caption{Pipeline of ZIP and UNZIP}
\label{corepipeline}
\end{figure}
\vspace{-1.5em}

The pipeline in Rocket Core is a 5-stage single-issue in-order pipeline. To reduce performance overhead, the calculations are processed in parallel. If the next instruction that uses a MAC calculation arrives after the previous one finished, the pipeline will not stall. Figure \ref{corepipeline} is a pipeline diagram of two instructions. In Figure \ref{corepipeline}, IF, ID, EX, MEM, WB stand for fetch, decode, execute, memory, and write back stages. UD/CK stands for updating Top register/checking MACs. As the figure shows, if the next ZIP/UNZIP instruction arrives after the MAC calculation, only one extra cycle is added to the pipeline, which is equivalent to inserting a nop. It is worth noting that the WB stage in ZIP/UNZIP does not rely on the finish of the UD/CK stage: in the write back stage, it only writes the \emph{ra} register, and the value does not rely on the current calculation. Fortunately, most functions require more cycles than 20. So in most cases, the ZIP/UNZIP instruction takes only one cycle. 

\subsubsection{Customized Compiler}

To make use of the new instructions, we also customized the riscv-gcc.  The modification on riscv-gcc is quite simple: Whenever we store a return address onto the stack, we add a ZIP instruction; whenever we pop a return address from the stack, we add a UNZIP instruction. It is noteworthy that, if a function will not call any function, \emph{ra} register will not be spilled to the stack. So we only add the new instructions when the \emph{ra} register is saved into/restored from the stack (rather than all calls and returns).



%

\subsubsection{Setjmp/Longjmp Support}

To support Setjmp/Longjmp, we also modified the glibc in the RISC-V tool chain. We have only modified two points:

\begin{enumerate}
\item Declaration of the Jump Buffer: Add additional space for the Top register and MAC.
\item Setjmp/Longjmp: Store/restore the Top register; Authenticate the data.
\end{enumerate}

Our changes perfectly support Setjmp/Longjmp, which is verified in some benchmarks in SPEC2000, such as perlbmk. These benchmarks will not pass without Setjmp/Longjmp support.

\subsubsection{Optimization}

In order to further reduce the runtime overhead, we also optimized the MAC module. We added a small cache (with a size of 4) to cache the recent results. If a new request can be found in the cache, the calculation will take only one cycle. This optimization slightly increased the complexity of the hardware, but significantly reduced runtime overhead (by around 30\%, see Section \ref{s_evaluation}). 


\subsubsection{A Comparable Hardware Based Shadow Stack}

In order to compare with Shadow Stacks, we also implemented a hardware supported Shadow Stack on Rocket Core. We tried to be consistent as much as possible: We added two instructions that back up or check the return address, and a pointer pointing the shadow stack. The compiler with Shadow Stacks inserts the instructions just like the way in Zipper Stack.


\subsection{LLVM}


In customized compiler approach, we implemented Zipper Stack algorithm based on LLVM 8.0. First we allocate some registers: we set the lower bits of XMM15 as the Top register, and several XMM registers as the Key register (we use different size of Key, corresponding to different numbers of rounds of AES-NI, see next part). We modified the backend of the LLVM so as to forbid these registers to be used by anything else. Therefore, the Key and Top register will not be leaked. At the beginning of a program, the Key and Top register are initialized by random numbers. We also modified the libraries so that the libraries will not use these registers.

\subsubsection{MAC function} Our implementation leverages the AES-NI instructions \cite{AES-NI} on the Inter x86-64 architecture to minimize the performance impact of MAC calculation. We use the Key register as the round key of AES-NI, and use one 128-bit AES block as the input (64-bit address and 64-bit MAC). The 128-bit result is truncated into 64-bit in order to fit our design. We use different rounds of AES-NI to test the performance overhead: a) standard AES, performs 10 rounds of an AES encryption flow, we mark it as \emph{full}; b) 5 rounds, marked as \emph{half}; c) one round, marked as \emph{single}. Obviously, the \emph{full} MAC function is the most secure one, and the \emph{single} is the fastest one.

\subsubsection{Operations} In each function, we insert a prologue at the entry, and an epilogue before the return. In the prologue, the old Top register is saved onto stack, and updated to the new MAC of the current return address and the old Top register value. In the epilogue, the MAC in the stack and return address are authenticated using the Top register. If it doesn't match, an exception will be raised. Just as we introduced before.




\subsection{Qemu}
To figure out whether it is possible to run the existing binaries if we change the logic of calls and returns, we customized the x86-64 instruction set and simulated it with Qemu. All the simulation is in the User Mode of Qemu 2.7.0.

The modification is quite concise: we add two registers and change the logic of Call and Ret instructions. As the algorithm designed, the Call instruction will push the address and the Top register, update the Top register with a new MAC, while Ret instruction will pop the return address and check the MAC. Since Qemu uses lower 39 bits to address the memory, we use the upper 25 bits to store the MAC. Correspondingly, the width of the Top register is also 25 bits. The Key register here is 64 bits. We used SHA-3 as the MAC function in this implementation. Since we did not change the stack structure, this implementation has good binary compatibility. Therefore it can help us to evaluate the security with real x86-64 attacks.

\section{Evaluation and Analysis}
\label{s_evaluation}

We evaluated Zipper Stack in two aspects: Runtime Performance and Security. We evaluated the performance overhead with the SPEC CPU 2000 on the FPGA and SPEC CPU 2006 in LLVM approach. 
Besides, we also tested the compatibility of modified x86-64 ISA with existing applications.

\begin{table}[thb]
  \caption{Security Comparison}
  \resizebox{\textwidth}{!}{
  \begin{tabular}{l|c|c|c|c|c}
   \toprule
   Adversary & Stack Protector & ASLR & Shadow Stack & MAC/Encryption & \textbf{Zipper Stack}\\
   \midrule
   Stack Overflow &safe&safe&safe&safe&safe\\
   \hline
   Arbitrary Write &unsafe&safe&safe&safe&safe\\
   \hline
   Memory Leak \& Arbitrary Write&unsafe&unsafe&unsafe&safe&safe\\
   \hline
   Secret Leak \& Arbitrary Write&N/A&N/A&N/A&unsafe&safe\\
   \hline
   Replay Attack &unsafe&unsafe&N/A&unsafe&safe\\
   \hline
   Brute-force Attempts &N/A&N/A&N/A&$2^{Ns-1}$&$2^{Ns-1}+N*2^{Nm-1}$ *\\
   \bottomrule
   \end{tabular}
   }
   \scriptsize{*: Under a probability of $1-(1-1/e)^N$, the valid collision of a certain attack does not exist.\\}
   \scriptsize{N: The number of gadgets in an attack; Nm: Bit width of MAC/ciphertext; Ns: Bit width of secret.\\}
   \scriptsize{The Brute-force attack already contains the memory leak, so the ALSR is noneffective and not considered.}
\label{sec}
\end{table}

\subsection{Security}

We first analyze the security of Zipper Stack, then show the attack test results, and finally, compare the security of different implementations.

\subsubsection{Security Analysis}

The challenge for the attacker is clear: how to tamper with the memory to make the fake return address be used and bypass our check? We list the defense effect of different methods of protecting return addresses in the face of different attackers in Table \ref{sec}. The table shows that Zipper Stack has higher security than Shadow Stack and other cryptography-based protection mechanisms.

\begin{figure}[hbt]
\begin{center}
\includegraphics[width=0.45\columnwidth]{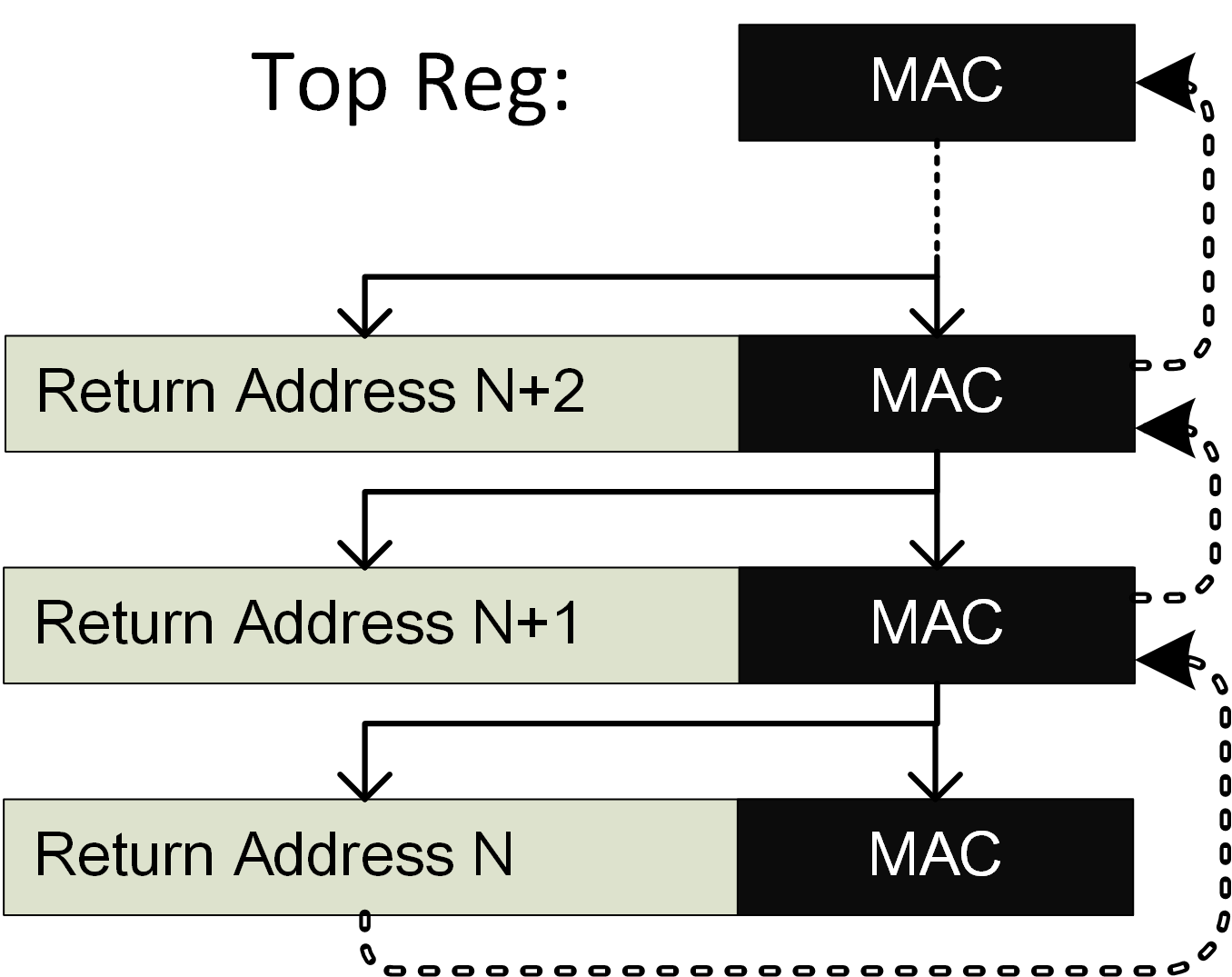}
\end{center}
\caption{\small{Direct Overwrite Attack: The solid lines indicate the protect relation of the MACs, and the dotted lines show the order that the attacker should overwrite in.}}
\label{attack}
\vspace{-2em}
\end{figure}

\emph{Direct Overwrite} First, we consider direct overwrite attacks. In the previous cryptography-based methods, the adversary cannot know the key and calculate the correct MAC, so it is secure. But we go further that the adversary may steal the key and calculate the correct MAC. As Figure \ref{attack} shows, to tamper with any return address structure and bypass the check (let's say, Return Address N), the attacker must bypass the pre-use check. Even if the attacker has stolen the key, the attacker needs to tamper with the MAC, which is used to check the return address, i.e., the MAC stored beside Return Address N+1. Since the MAC and the return address are authenticated together, the attacker has to modify the MAC stored beside Return Address N+2 to tamper with the MAC bound to Return Address N+1. And so on, the attacker has to alter the MAC at the top, which is in the Top register. As we have assumed, the register is secure against tampering. As a result, an overwrite attack won't work.

\emph{Replay Attack} Next, we discuss replay attacks: since we have assumed that the attacker can read all the memory, is the attacker able to utilize the protected addresses and their MACs in memory to play a replay attack? In Zipper Stack, it is not feasible because once the call path has changed, the MAC will be updated immediately. The MACs in an old call path cannot work in a new call path, even if the address is the same. This is an advantage over previous cryptography-based methods: Zipper Stack resist replay attacks naturally.

\emph{Brute-force Attack} Then we discuss the security of Zipper Stack in the face of brute force. Here we consider the attacks which read all related data in memory, guess the secret constantly, and finally construct the attack. In the cryptography-based approaches, security is closely related to the entropy of the secret. Here in Zipper Stack, the entropy of the secret is the bit width of Key register (Ns), which means the attacker needs to guess the correct Key register in a space of size $2^{Ns}$. Once the attacker gets the correct key, the attack becomes equivalent to an attack with a secret leak.

\emph{Attack with Secret Leak} The difference between Zipper Stack and other cryptography-based approaches is, other approaches will fail to protect control flow once the attacker knows the secret, but Zipper Stack will not. Because even if the attacker knows the secret key, the Top register cannot be altered. If an attack contains N gadgets, the attacker needs to find N collisions whose input contains the ROP (or ret2lib) gadget addresses to use the gadgets and bypass the check\footnote{The same gadget addresses won't share the same collision, because the MACs bound to them differ.}. Considering an ideal MAC function, one collision will take about $2^{Nm-1}$ times of guesses on average\footnote{The MAC function is a $(Nm+Na)$bit-to-$Nm$bit function, so choosing a gadget address will determine Na bits of input, which means there are $2^{Nm}$ optional values. On average, there is one collision in the values, since every $2^{Nm+Na}/2^{Nm}=2^{Na}$ inputs share the same MAC value. Because of our special algorithm, this is not a birthday attack nor an ordinary second preimage attack, but a limited second preimage attack.}. So an attack with N gadgets will take $N*2^{Nm-1}$ times of guesses on average even if the Key register is leaked. The total number of guesses is (guessing Key value and the collisions) $2^{Ns-1}+N*2^{Nm-1}$. And more unfortunate for the attackers: under a certain probability ($1-(1-1/e)^N$ for an attack contains N gadgets), the valid collision does not even exist. Take an attack that contains five gadgets as an example, the possibility that the collision does not exist is around 90\%, and the possibility grows as the N grows.

\subsubsection{Attack Tests} We tested some vulnerabilities and the corresponding attacks to evaluate the security of Zipper Stack. In these tests, we use Qemu implementation, because most attacks are very sensitive to the stack layout, a customized compiler (or just a compiler in a different version) may lead to failures. Using Qemu simulation can keep the stack layout unchanged, avoid the illusion that the defense works, which is actually because of the stack layout changes.  We wrote a test suite contains 18 attacks and the corresponding vulnerable programs
. Each attack contains at least one exploit on return addresses, including stack overflow, ROP gadget, or ret2lib gadget. All attacks are detected and stopped (all of them will alter the MAC and cannot pass the check during return). These tests show that Zipper Stack is reliable. Table \ref{t_attack} shows the difference between Zipper Stack and other defense methods.
\vspace{-1.5em}
\begin{table}[thb]
  \caption{Attack Test against Defenses}
  \label{t_attack}
  \begin{center}
  \begin{tabular}{l|r|r|r}
  \toprule
  Defence            & \# Applied & \# Secured & \# Bypassed \\ \midrule
  DEP                & 20      & 2       & 18        \\ \hline
  ASLR               & 8       & 2       & 6        \\ \hline
  Stack Canary       & 6       & 2       & 4        \\ \hline
  Coarse-Grained CFI & 16      & 10      & 6        \\ \hline
  \textbf{Shadow Stacks}      & 2       & 0       & 2        \\ \hline
  \textbf{Zipper Stack}       & 20      & 20      & 0      \\ \bottomrule
  \end{tabular}
  \end{center}
\end{table}

\vspace{-2em}

\emph{Attacks Against Shadow Stacks} The above attacks can also be stopped by Shadow Stacks. To further prove the Zipper Stack's security advantages compared to Shadow Stacks, we also wrote two proof-of-concept attacks (corresponding to two common types of shadow stack) that can bypass the shadow stack. As introduced in previous work \cite{losing}, Shadow Stack implementations have various flaws and can be attacked via different vulnerabilities. We constructed similar attacks. In both PoC attacks, we corrupt the shadow stack before we overwrite the main stack and perform ROP attacks. In the first example, the shadow stack is parallel to the regular stack, as introduced in \cite{Cost-of-SS:2015}. The layout of the shadow stack is easy to obtain because its offset to the main stack is fixed. In the second example, the shadow stack is compact (only stores the return address). The offset of this shadow stack to the main stack is not fixed, so we used a memory leak vulnerability to locate the shadow stack and the current offset. In both cases, Shadow Stack can not stop the attacks, but Zipper Stack can. We also added both attacks to Table \ref{t_attack}.

\subsubsection{Security of Different Approaches}

The security of Zipper Stack is mainly affected by two aspects: the security of Top register and the length of MAC, corresponding to the risk of tamper of MAC and the risk of brute-force attack. From the Top register's perspective, the protection in the hardware and Qemu approach is complete. In both implementations, the program cannot access the Top register except the \emph{call} and \emph{return}, so there is no risk of tampering. In the LLVM approach, an XMM register is used as the Top register, so it faces the possibility of being accessed. We banned the use of XMM15 in the compiler and recompiled the libraries to prevent access to the Top register, but there are still risks. For example, there may be unintended instructions in the program that can access the XMM15 register. From the perspective of MAC length, MAC in LLVM approach is 64-bit wide and therefore has the highest security. MAC in hardware is 24-bit wide and Qemu is 25-bit wide, so they have comparable security but still lower than LLVM approach.


\subsection{Performance Overhead}

We evaluated the performance overhead of different approaches. The cost of the hardware approach is significantly lower than that of the LLVM approach. It is worth noting that QEMU is not designed to reflect the guest's actual performance, it only guarantees the correctness of logic \footnote{For example, QEMU will optimize basic blocks. Some redundant instructions (e.g. a long nop slide) may be eliminated directly, which cannot reflect real execution performance. }. Thus, we do not evaluate the performance overhead of the QEMU approach.

\subsubsection{SPEC CINT 2000 in Hardware Approach}

\begin{table*}[t]
\caption{Result of SPEC 2000 in Hardware Approach}
\label{fpga}
\begin{center}
\resizebox{0.9\textwidth}{!}{
\begin{tabular}{l|rrrr}
 \toprule
             & Baseline  & Shadow Stack & Zipper Stack & Zipper Stack (optimized) \\
 Benchmark   & (seconds) & (seconds/slowdown)& (seconds/slowdown)& (seconds/slowdown)\\
 \hline
164.gzip & 10923.10  & 10961.65   ( 0.35\% )& 10960.60   ( 0.34\% )& 10948.88  ( 0.24\% )\\
175.vpr & 7442.48  & 7528.06   ( 1.15\% )& 7490.49   ( 0.65\% )& 7485.40  ( 0.58\% )\\
176.gcc & 8227.93  & 8318.83   ( 1.10\% )& 8348.99   ( 1.47\% )& 8317.34  ( 1.09\% )\\
181.mcf & 11128.67  & 11153.01   ( 0.22\% )& 11183.31   ( 0.49\% )& 11168.93  ( 0.36\% )\\
186.crafty & 10574.27  & 10942.89   ( 3.49\% )& 10689.74   ( 1.09\% )& 10692.53  ( 1.12\% )\\
197.parser & 8318.16  & 8577.67   ( 3.12\% )& 8658.89   ( 4.10\% )& 8544.72  ( 2.72\% )\\
252.eon & 14467.81  & 15111.99   ( 4.45\% )& 15519.98   ( 7.27\% )& 15040.26  ( 3.96\% )\\
253.perlbmk & 7058.96  & 7310.78   ( 3.57\% )& 7388.20   ( 4.66\% )& 7342.20  ( 4.01\% )\\
254.gap & 7728.56  & 7850.32   ( 1.58\% )& 7926.10   ( 2.56\% )& 7817.52  ( 1.15\% )\\
255.vortex & 13753.47  & 14738.06   ( 7.16\% )& 14748.70   ( 7.24\% )& 14644.70  ( 6.48\% )\\
256.bzip2 & 6829.01  & 6893.50   ( 0.94\% )& 6954.81   ( 1.84\% )& 6865.27  ( 0.53\% )\\
300.twolf & 11904.25  & 12044.16   ( 1.18\% )& 11974.22   ( 0.59\% )& 11917.37  ( 0.11\% )\\

 \hline
Average&&2.36\%&2.69\%&1.86\%\\
 \bottomrule
\end{tabular}}
\end{center}
\vspace{-2em}
\end{table*}

To evaluate the performance of Zipper Stack on RISC-V, we instantiated it on the Xilinx Zynq VC707 evaluation board and ran the SPEC CINT 2000 \cite{SPEC:2000} benchmark suite (due to the limited computing power of the Rocket Chip on FPGA, we chose the SPEC 2000 instead of SPEC 2006). The OS kernel is Linux 4.15.0 with support for the RISC-V architecture. The hardware and GNU tool-chain are based on \emph{freedom} (commit $cd9a525$) \cite{freedom}. All the benchmarks are compiled with GCC version 7.2.0 and -O2 optimization level. We ran each benchmark for 3 times.

Table \ref{fpga} shows the results of Zipper Stack and Shadow Stacks.
The result shows that without optimization, Zipper Stack is slightly slower than Shadow Stacks (2.69\% vs 2.36\%); while with optimization (the cache), Zipper Stack is much faster than Shadow Stacks (1.86\% vs 2.36\%). To sum up, the runtime overhead of Zipper Stack is satisfactory (1.86\%).

\subsubsection{Performance Overhead in LLVM Approach}

\begin{table*}[hbt]
  \vspace{-3em}
  \caption{Result of SPEC 2006 in LLVM approach}
  \begin{center}
  \resizebox{\textwidth}{!}{
  \begin{tabular}{l|rrrrr}
   \toprule
    &Baseline&Full&Half&Single\\
   Benchmark   & (seconds) & (seconds/slowdown)& (seconds/slowdown)& (seconds/slowdown)\\
   \hline
   401.bzip2 & 337.69  & 385.98  ( 14.30\% )& 358.32  ( 5.76\% )& 336.33  ( -0.40\% )\\
   403.gcc & 240.38  & 341.06  ( 41.89\% )& 305.94  ( 21.43\% )& 272.42  ( 13.33\% )\\
   445.gobmk & 340.28  & 493.47  ( 45.02\% )& 427.90  ( 20.48\% )& 384.91  ( 13.12\% )\\
   456.hmmer & 269.62  & 270.50  ( 0.33\% )& 268.81  ( -0.30\% )& 268.74  ( -0.33\% )\\
   458.sjeng & 403.38  & 460.58  ( 14.18\% )& 435.94  ( 7.47\% )& 421.69  ( 4.54\% )\\
   462.libquantum & 251.50  & 265.50  ( 5.57\% )& 261.17  ( 3.70\% )& 256.72  ( 2.07\% )\\
   464.h264ref & 338.29  & 410.56  ( 21.37\% )& 381.11  ( 11.24\% )& 355.97  ( 5.23\% )\\
   473.astar & 285.58  & 362.00  ( 50.11\% )& 333.91  ( 14.47\% )& 311.26  ( 8.99\% )\\
   \hline
   \textbf{Average} &  &  21.13\% &  8.96\% &  4.75\% \\
   \bottomrule
\end{tabular}}
\end{center}
\vspace{-1.5em}
\label{llvm}
\end{table*}

To evaluate the performance of Zipper Stack on customized compiler, we run the SPEC CPU 2006 \cite{SPEC:2006} compiled by our customized LLVM (with optimization -O2). 

Table \ref{llvm} shows the performance overhead of Zipper Stack in LLVM approach\footnote{The performance gain is due to memory caching artifacts and fluctuations.}. Shadow Stack is reported to cost about 2.5-5\% \cite{SoKWar,Cost-of-SS:2015}.
Our approach costs 4.75\% $\sim$ 21.13\% accroding to the results, depending on how many rounds we perform in MAC function. It does not cost too much overhead if we use only one round of AES-NI, although it is still slower than hardware approach. Furthermore, it costs 21.13\% when we perform a standard AES encryption, which is faster than CCFI \cite{CCFI:2015}, since we encrypt less pointers than CCFI.







\subsection{Compatibility Test}

Here, we test the binary compatibility of Zipper Stack. This test is only valid for Qemu implementation. In the other two implementations, due to we have modified the compiler, we can use Zipper Stack as long as we recompile the source code, so there is no compatibility issue. The purpose of this test is: If we only modify the Call and Ret instructions in the x86-64 ISA, and use the compression structure to maintain the stack layout, is it possible to maintain binary compatibility directly?

We randomly chose 50 programs in Ubuntu (under the path /usr/bin) to test the compatibility in Qemu. 42 out of 50 programs are compatible with our mechanism.  Most failures are due to the Setjmp/Longjmp, which we have not supported in Qemu yet. So we think although some issues need to be solved (such as the setjmp/longjump), Zipper Stack can be used directly on most existing x86-64 binaries.

\section{Conclusion}
\label{s_conclusion}

In this paper, we proposed Zipper Stack, a novel algorithm of return address protection, which authenticates all return addresses by a chain structure using MAC. It minimizes the amount of state requiring direct protection and costs low performance overhead.

Through our analysis, Zipper Stack is an ideal way to protect return addresses, and we think it is a better alternative to Shadow Stack. We discussed various possible attackers and attacks in detail, concluding that an attacker cannot bypass Zipper Stack and then counterfeit the return addresses. In most cases, Zipper Stack is more secure than existing methods. The simulation of attacks on Qemu also corroborates the security of Zipper Stack. Our experiment also evaluated the runtime performance of Zipper Stack, and the results have shown that the performance loss of Zipper Stack is very low. The performance overhead with hardware support based on Rocket Core is only 1.86\% on average (versus a hardware-based Shadow Stack costs 2.36\%). 
Thus, the proposed design is suitable for actual deployment.




\bibliographystyle{splncs04}
\bibliography{readm}

\end{document}